 \documentclass[final,5p,times,twocolumn,authoryear]{elsarticle}

\newcommand{\mescal}{\textsc{mescal}}

\usepackage{amssymb}
\usepackage{amsmath}
\usepackage{lipsum}
\usepackage{rotating}
\usepackage{url}
\usepackage[T1]{fontenc}
\usepackage[utf8]{inputenc}
\usepackage{color}

\journal{Astronomy $\&$ Computing}

\begin{document}

\begin{frontmatter}


\title{The HERMES Calibration Pipeline: \mescal}


\author[iaps]{G.~Dilillo}

\author[oas]{E.~J.~Marchesini}

\author[iaps]{G.~Della~Casa}

\author[oa]{G.~Baroni}

\author[oas]{R.~Campana}

\author[oas]{E.~Borciani}

\author[oas]{S.~Srivastava}

\author[oa]{S~Trevisan}

\author[iaps]{F.~Ceraudo}

\author[oa]{M.~Citossi}

\author[iaps]{Y.~Evangelista}

\author[tub]{A.~Guzm\'an}

\author[tub]{P.~Hedderman}

\author[oas]{C.~Labanti}

\author[oas]{E.~Virgilli}

\author[oa]{F.~Fiore}

\affiliation[iaps]{organization={INAF-IAPS},
            addressline={Via del Fosso del Cavaliere 100}, 
            city={Rome},
            postcode={I-00133}, 
            country={Italy}
            }

\affiliation[oas]{organization={INAF-OAS},
            addressline={Via Piero Gobetti 101}, 
            city={Bologna},
            postcode={I-40129}, 
            country={Italy}
            }

\affiliation[oa]{organization={INAF-OATS},
            addressline={Via Giambattista Tiepolo 11}, 
            city={Trieste},
            postcode={I-34131}, 
            country={Italy}
            }

\affiliation[tub]{organization={IAAT-Universit\"at Tübingen},
            addressline={Geschwister-Scholl-Platz}, 
            city={T\"ubingen},
            postcode={D-72074}, 
            country={Germany}
            }

\begin{abstract}
The HERMES Technologic and Scientific Pathfinder project is a constellation of six CubeSats aiming to observe transient high-energy events such as the Gamma Ray Bursts (GRBs). HERMES will be the first space telescope to include a \emph{siswich} detector, able to perform spectroscopy in the 2 keV to 2 MeV energy band. The particular siswich architecture, which combines a solid-state Silicon Drift Detector and a scintillator crystal, requires specific calibration procedures that have not been yet standardized in a pipeline.  We present in this paper the HERMES calibration pipeline, \mescal, intended for raw HERMES data energy calibration and formatting. The software is designed to deal with the particularities of the siswich architecture and to minimize user interaction, including also an automated calibration line identification procedure, and an independent calibration of each detector pixel, in its two different operating modes. The \mescal\ pipeline can set the basis for similar applications in future siswich telescopes. 
\end{abstract}




\end{frontmatter}
\tableofcontents

\section{The HERMES Pathfinder project}
\label{sec1}

The \emph{High Energy Rapid Modular Ensemble of Satellites} (HERMES)\footnote{\url{https://www.hermes-sp.eu}} Technologic and Scientific Pathfinder project will be the first high-energy transient localization experiment through a distributed space architecture realized with a 3+3 CubeSat constellation \citep{HERMES}. Each unit will be equipped with a new miniaturized instrument, hosting a hybrid Silicon Drift Detector (SDD) and a cerium-doped gadolinium-aluminium-gallium
garnet (GAGG:Ce) scintillator photodetector system. This complex photodetector is sensitive to both X-rays and $\gamma$-rays \citep{evangelista22}. These detectors, exploiting the so-called \emph{siswich} architecture \citep{marisaldi06}, aim to monitor the sky looking for high-energy transients, such as Gamma Ray Bursts and the electromagnetic counterparts of Gravitational Wave events.

The first of its kind, the HERMES modular architecture will provide accurate positioning of the detected sources by exploiting its timing capabilities, through the triangulation technique \citep{sanna20}. Thus, HERMES will be able to provide a fast and inexpensive complement to more ambitious missions, with a cost one order of magnitude less than conventional high-energy observatories.

\begin{figure}
    \centering
    \includegraphics[width=0.3\textwidth]{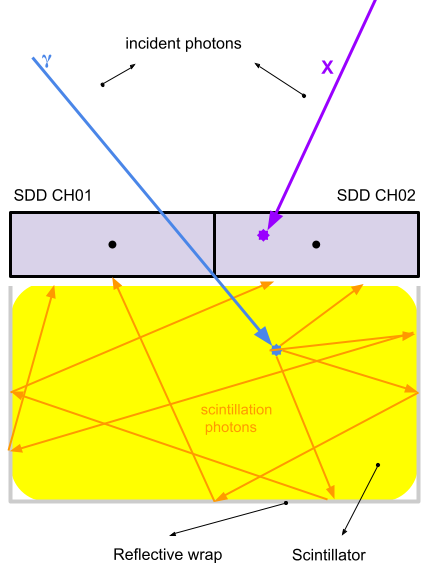}\\
    \caption{An example of the working principle for a siswich detector in the HERMES architecture. Single incident X-ray photons are detected directly by an individual SDD cell (X-mode). Individual incident higher-energy $\gamma$-ray photons pass through the SDDs and are absorbed into the wrapped scintillator crystal, which re-emits scintillation optical/UV light. The scintillation photons are then collected by a pair of SDDs optically coupled with the crystal (S-mode). The output signals collected by the FEE of this pair of channels are then compared to discriminate between S-mode and X-mode events.}
    \label{siswich}
\end{figure}

While most astronomical detectors work with only one detection principle, the siswich detectors exploit two: the SDD as a direct X-ray photon detector, and the scintillator crystal as an indirect $\gamma$-ray detector, both working as a monolithic, single-acquisition instrument (see Figure~\ref{siswich}). This ``double detection'' mechanism is the key to achieving the broad energy range of sensitivity of HERMES, which spans approximately from $\sim$2 keV to 2 MeV \citep[see ][]{campana20,fuschino20,evangelista22}. 
However, this mechanism also poses challenges when it comes to spectroscopic data calibration. To be calibrated, a payload is exposed to multiple radioactive sources and events observed by different instrument's channels are classified according to their source and detection principle. This operation allows for the measurement of the instrument's parameters, which can later be used to measure the energy of photons from arbitrary sources. Each HERMES payload hosts 120 channel, each effectively working as an independent detector and requiring characterization. During the assembly of a single payload, the calibration procedure is repeated several times, both to ensure the success of each integration step and to comprehensively characterize the payload under different operational conditions. Given the number of operations involved, some level of software automation is practically indispensable for calibrating the HERMES payloads

In this paper, we introduce \mescal\ (\emph{herMES CALibration} pipeline), a software tool that implements a pipeline for calibrating siswich detectors. While \mescal\ has been tailored for the unique specifications of the HERMES detector, the workflow it implements and documents can be adapted to different instruments with minimal effort. In this sense, \mescal\ aims to provide a flexible and standardized framework for the calibration of siswich detectors with different designs.

This article is organized as follows: in Section \ref{sec2} we discuss the HERMES working principle and raw data format, in Section \ref{sec3} we discuss the general structure of the \mescal\ pipeline and its data products, in Section \ref{sec4} we detail some of the algorithms implemented within \mescal, and in Section \ref{conc} we summarize our conclusions.

\section{The HERMES data format and requirements}
\label{sec2}

\subsection{Detector working principle}
\label{subsec21}

\begin{figure}
    \centering
    \includegraphics[width=0.5\textwidth]{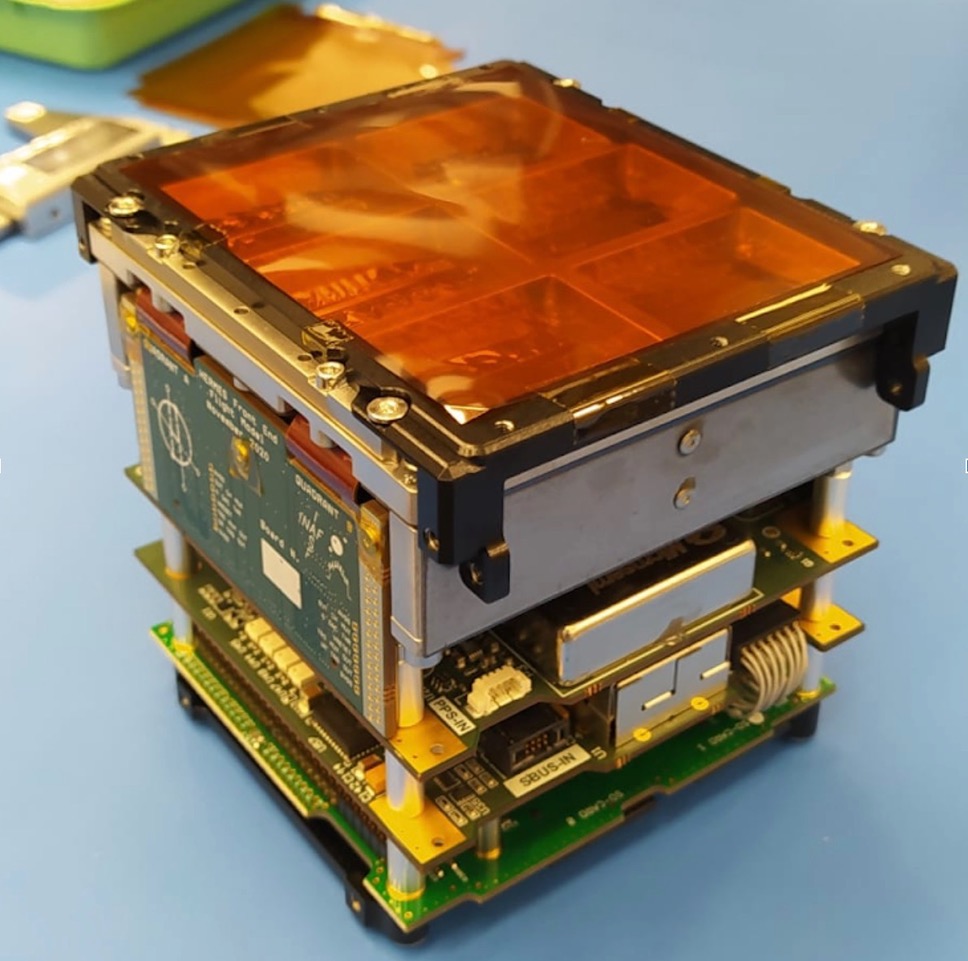}\\
    \caption{An individual HERMES payload flight model. Its dimensions are $10\times10\times10$~cm$^3$.}
    \label{hermes}
\end{figure}

A single HERMES flight model detector (Figure~\ref{hermes}) consists of four electrically-independent quadrants, for a total of 120 readout channels, corresponding to 120 SDD cells, each with a $\sim$45 mm$^2$ sensitive area and split in 12 monolithic 2$\times$5 matrices. The SDDs are, in turn, optically coupled to scintillator crystals in pairs (see Figure \ref{siswich}). This design choice allows to discriminate between incoming X-ray ($\sim$3 to 60 keV) and $\gamma$-ray ($\sim$ 20 to 2000 keV) photons. The former are detected directly by the SDDs, while the latter deposit their energy in a GAGG:Ce scintillator crystal, which then re-emits a proportional quantity of optical photons, read-out by the same SDDs. Since each scintillator crystal is optically coupled to two SDDs, when a simultaneous event is read by both SDDs, the event is classified as a $\gamma$-ray event. 

The detector has therefore a different response to direct X-ray absorption (the so-called X-mode) and to (indirect) $\gamma$-ray detection (S-mode), and shall be separately calibrated with appropriate functions for each operating mode.

The SDD detectors are read-out by an Application Specific Integrated Circuit (ASIC) named LYRA \citep{fuschino20,grassi20,gandola21}, which allows for a low-noise, low-power electronic front-end, and which provides the necessary shaping and discrimination logic. The analog-to-digital conversion and timestamping of the event is performed in the back-end electronics (BEE) board \citep{evangelista22}, with an on-board Virtex-5 FPGA. A commercial payload data handling unit (PDHU, \citealt{guzman20}) handles the communication with the BEE on one side and the spacecraft bus on the other side, the housekeeping data and the preparation of event packets. In Table \ref{reqhermes} we list the most important requirements for the HERMES mission, regarding spectroscopy.

\begin{table*}
\begin{center}
\caption{Requirements regarding spectroscopy performance for the HERMES-Pathfinder mission. Adapted from \citet{evangelista22}.}
\begin{tabular}{ c|c }
Requirement & Condition  \\
\hline
Sensitivity & $\leq$ 2 ph/s/cm$^2$ (E$\leq$20 keV)\\ 
 & $\leq$ 1 ph/s/cm$^2$ (50 keV $\leq$ E $\leq$ 300 keV)  \\ 
Energy band & 5 keV $\leq$ E $\leq$ 500 keV  \\ 
Energy resolution EOL & $\leq$1 keV FWHM (5.0 to 6.0 keV)\\ 
 & $\leq$5 keV FWHM (50.0 to 60.0 keV)  \\ 
\end{tabular}
\label{reqhermes}
\end{center}
\end{table*}

\subsection{Raw data format}
\label{subsec22}
Event lists are provided by the BEE once per second, encapsulated with housekeeping data within the PDHU and sent as scientific telemetry packets by the spacecraft. 
When a trigger occurs (i.e., when one or more detector channels detect one event above a programmable threshold) the BEE commands the digital-to-analog conversion of all the above-threshold channels in the given detector quadrant, timestamps the events and provides them to the PDHU. The event list is formatted as a stream of 64-bytes words, which could contain timing information or the address and amplitude of the triggering channel. The timing information is derived from a 100~ns counter, reset at the pulse-per-second (PPS) signal from the on-board GPS receiver. An ultra-stable chip-scale atomic clock (CSAC) guarantees the clock synchronization and thus time-stamping accuracy also when the GPS is not locked.

The raw telemetry data format is then converted in FITS files on ground, with the appropriate structure (Level 0, with all the uncalibrated raw data) and processed by the standard scientific pipeline (producing Level 1 calibrated data, and Level 2 scientific products).

\section{The \mescal\ pipeline}
\label{sec3}

The \mescal\ pipeline has been developed during the HERMES integration and testing activities, to allow for a simple and immediate reduction of the data acquired during the long and complex calibration activities \citep{campana22}. 

It is worth noting that \mescal, although similar in structure, will not be part of the standard scientific calibration pipeline (which is being designed using HEASARC\footnote{The High Energy Astrophysics Science Archive Research Centre, managed by NASA Goddard Space Flight Center, handles several standard multi-mission data exploration and analysis tools, see e.g., \url{https://heasarc.gsfc.nasa.gov/docs/software.html}} compatible tools at SSDC\footnote{ASI Science Data Center, \url{https://www.ssdc.asi.it/}}). Instead, \mescal\ has been designed to analyze a large amount of raw data, aimed at obtaining the best calibration parameters that will later on be included in the calibration database. In particular, this calibration database will be then accessed by the standard scientific pipeline to obtain the instrumental parameters necessary to reduce scientific data, under different in-orbit conditions. Thus, while the scientific pipeline HEASARC-based tools implement a calibration with known parameters, \mescal\ is designed to find these parameters from raw laboratory data,  in the first place, which will then used as a reference for the scientific calibration of all flight data.

Usually, the instrument calibration in both its operating modes (X and S-mode) is performed by means of suitable radioactive sources, which emits X and $\gamma$-rays of known energies. Sources usually employed are, for example, $^{55}$Fe (lines at 5.89 and 6.49 keV), $^{109}$Cd (lines at 21 and 25 keV) or $^{137}$Cs (monochromatic $\gamma$-ray photon at 662 keV).

\subsection{General procedure}
\label{subsec31}

The \mescal\ pipeline is implemented in Python 3.10 and provides a text-based user interface.  
It requires a Level 0.5 data file (i.e., a modified FITS Level 0 file), and to specify the Flight Model to be calibrated, as well as the calibration sources that were used during the acquisition. The Level 0.5 data file differs from the Level 0 file mainly in that the amplitude and address information is stored in fixed, separate FITS table columns, to ease the quicklook and access. The FITS Level 0 file, instead, for storage optimisation saves these informations in variable-length arrays.

The associated decay energies are stored in one of the support libraries within \mescal. Handling of the FITS files is done with the \texttt{astropy} package within Python \citep{astropy}, version 5.1.
During initialization, \mescal\ reads both the detector map corresponding to the specific flight model (containing the raw address to specific SDD/crystal ID conversion information, which differs between flight models), and the data file. All the relevant information from the latter is stored in a data table (a \texttt{pandas} dataframe table, \citet{reback2020pandas}). The dataset is filtered from events due to electronic noise. These events are excluded from the calibration process but retained in a dedicated table to enable supplementary analysis, if needed. Events are then classified into X and S-mode events, using the appropriate detector's pixel map, and a mode tag column is added to the data table.\\
Once tagged, the calibration of the X-tagged events begins. First, a collection of spectra is generated for these events, by building histograms of their amplitude in instrumental units, separately for each channel.
A peak detection algorithm searches for a set of local maxima (i.e., line peaks), given by the number of calibration lines to use, within each channel's X histogram. The working principle of this algorithm is described in Sect. \ref{subsec42}. Once the algorithm selects a set of local maxima, it defines a suitable range around each maximum, spanning the line peaks. A Gaussian profile is fitted to these local maxima through least-square minimization. The fit results determine the position and width of the X emission lines for each channel in instrumental units.
Through a linear regression of the known line energies vs. the centroids of the detected peaks, the channels' \emph{gain} and \emph{offset} parameters are estimated:
\begin{equation}
    A_i [\mathrm{ADC}] = \mathrm{Gain} \cdot E_i [\mathrm{keV}] + \mathrm{Offset}
\end{equation}
Here, $A_i [ADC]$  represents the $i$th-centroids position in instrumental units and $E_i$ represent the emission line associated to it.
This step allows for the amplitude of the X-tagged events to be expressed in energy units, thereby completing their calibration process \citep{campana22}.
Following this step, the calibration of the S-tagged events begins.
Assuming that for each $3.65$ eV of incident radiation energy electron-hole pair are freed into the silicon \citep{mazziotta2008electron}, the measurement of the gain and offset parameters also enables the expression of the events' amplitude in units of photoelectrons. This fact is exploited to gauge the total number of photoelectrons freed over a channel couple by each scintillation event. The amplitude of the S-tagged events are converted to electrons, using the gain and offset parameters estimated from the X calibration of the appropriate channel, according to:
\begin{equation}
    A [\mathrm{e}^{-}] = \frac{A [\mathrm{ADC}] - \mathrm{Offset}}{\mathrm{Gain}}
\end{equation}
Then, a new event data table is aggregated summing the electron-amplitude of coincident S-events, i.e. S-tagged events which are simultaneous and took place over coupled channel pairs. A new collection of histograms is built starting from this data table, one histogram per scintillator (or equivalently, channel couple). These histograms are searched for emission peaks due to the $\gamma$ emission lines, which are subsequently fitted to gaussian profiles. The best fit parameters allows the measure of each scintillators' effective light output, according to:
\begin{equation}
\label{scalib1}
LY[\mathrm{e}^-/\mathrm{keV}]=\frac{A [\mathrm{e}^{-}]}{A[\mathrm{keV}]}
\end{equation}
where $LY$ is the effective light output for the scintillator, $A [\mathrm{e}^-]$ is the fit's centroid position in electron units, and $A [\mathrm{keV}]$ is the energy of a calibration line used as reference, usually the $661.67$~keV line of $^{137}$Cs. 
Moreover, an \emph{effective light-output} value is also assigned to each channel through:
\begin{equation}
\frac{LY_1}{LY} = \frac{A_1}{A}
\quad\mathrm{and}\quad 
\frac{LY_2}{LY} = \frac{A_2}{A}
\end{equation}
Where $LY$ is the light-yield computed for the channel couple's scintillator, $A_1$ and $A_2$ are the centroid of the calibration line in the individual channels' spectra, and $LY_1$ and $LY_2$ are the effective light output of the two, coupled channels.
Once the effective light output has been computed for each channel, the amplitude of the aggregated S-tagged events can be converted from photoelectron units to energy. The calibrated X-tagged events and the calibrated S-tagged events are merged together in a calibrated event list which is the final output of the calibration procedure. \\

\subsection{Data products}
\label{subsec32}

One of the main requirements for the \mescal\ pipeline is to be able to produce also data products that are easy to analyze, in order to quickly diagnose the overall performance of the detector during the integration and calibration activities. Thus, the final data products of the \mescal\ pipeline are classified into three main categories: \emph{results tables} (or \emph{reports}), \emph{data plots}, and an \emph{event list}. The tables should contain all relevant information in detail, while the plots are designed to provide a quicklook on some of the most important performance aspects. Plots within \mescal\ are built using the \texttt{matplotlib} package within Python \citep{hunter07}, version 3.6.

The event list consists of an energy-calibrated table of photon events. This list is exported in the FITS format and contains all events amplitudes, both in energy and in electrical charge units, divided by type (X or S), with their respective pixel addresses (CH and QUADID), and time tag. Time units are instrumental, i.e., seconds from the start of the last data acquisition, while in flight time units will instead be expressed as Mission Elapsed Time (MET), i.e., elapsed seconds from January 1st, 2022 00:00 in Terrestrial Time (TT).

\subsubsection{Data reports}
\label{subsec321}

The \mescal\ data reports are designed to contain all the relevant information for calibration purposes. 
Each step of the calibration process can be assessed by the results listed on these reports. The reports are presented in either FITS, XLSX or CSV formats, as chosen by the user. The \texttt{xfit} and \texttt{sfit} reports include all the parameters used during the fitting of the spectral features (see Table \ref{xfit} and Table \ref{sfit}), while the \texttt{cal} and \texttt{slo} reports show the gain and offset, and the effective light output, respectively (see Tables \ref{cal} and \ref{slo}). The \texttt{res} report shows the resolution of the X-mode emission features, defined as the full width at half maximum of their Gaussian profiles (see Table \ref{res}).
As such, the \mescal\ outputs provide the necessary input to synthesize the final calibration files (stored in a standard calibration database, CALDB) used in the standard scientific pipeline.
The CALDB structure revolves around several file formats, containing the needed information to be readout by the standard pipeline during the scientific data reduction. For example, the pipeline will read out gain and offset values from a opportunely structured FITS file, interpolate them to the actual observation temperature and use this information to equalize all the detector channels. Other CALDB files encode informations about the active channels, look-up tables, FEE configuration histories and so on.

    \begin{table*}
        \caption[]{Example of a report on the X-mode fit results. It consists of an array of tables, one per each calibration source. The table displays the corresponding channel in column 1, the centroid of the fit of the line and its error in columns 2 and 3, the FWHM of the fitted Gaussian profile and its error in columns 4 and 5, the amplitude of the Gaussian profile and its error in columns 6 and 7, and the low and high limits of the range within which the algorithm concluded there is a line to fit, in columns 8 and 9. The table can eventually continue, with extra columns reporting the same information but for a different emission line.}
        \label{xfit}
        \centering
        \begin{tabular}{|c|c|c|c|c|c|c|c|c|c|c|}
            \hline
            source & \multicolumn{8}{c|}{Fe-55 (5.9 keV)}  & \multicolumn{2}{c}{Cd-109 (22.1 keV) ... } \\
            parameter & center & center err & fwhm & fwhm err & amp & amp err & lim low & \multicolumn{1}{c|}{lim high} & center & \multicolumn{1}{c}{...}\\
            \hline
            channel			& & & & & & & & \multicolumn{1}{c|}{}& & \multicolumn{1}{c}{...}\\
            0 & 16883 & 0.5 & 45.1 & 1.0 & 46600 & 1014 & 16844 & \multicolumn{1}{c|}{16901} & 19321 \\
            1 & 16517 & 0.5 & 62.7 & 0.9 & 43238 & 580 & 16464 & \multicolumn{1}{c|}{16542} & 19411 \\

        \end{tabular}
    \end{table*}

    \begin{table*}
        \caption[]{Example of a report on the S-mode fit results. The table displays the corresponding channel in column 1, the centroid of the fit of the line and its error in columns 2 and 3, the FWHM of the fitted Gaussian profile and its error in columns 4 and 5, the amplitude of the Gaussian profile and its error in columns 6 and 7, and the low and high limits of the range within which the algorithm concluded there is a line to fit, in columns 8 and 9. The table can eventually continue, with extra columns reporting the same information but for a different emission line, if any.}
        \label{sfit}
    \centering
        \begin{tabular}{|c|c|c|c|c|c|c|c|c|}
            \hline
            source & \multicolumn{8}{c|}{Cs-137 (661.7 keV)}  \\
            parameter & center & center err & fwhm & fwhm err & amp & amp err & lim low & lim high \\
            \hline
            channel			& & & & & & & & \\
            0 & 21279 & 17 & 676 & 34 & 17852 & 760 & 21000 & 21838 \\
            1 & 21995 & 14 & 684 & 27 & 17865 & 644 & 21726 & 22571 \\
        \end{tabular}
    \end{table*}

    \begin{table*}
        \caption[]{Example of a report on the X-mode calibration results. The table displays the corresponding channel in column 1, the gain and its error in columns 2 and 3, and the offset and its error in columns 4 and 5. The Pearson correlation coefficient $r$ is displayed in column 6.}
        \label{cal}
    \centering
        \begin{tabular}{|c|c|c|c|c|c|}
            \hline
            channel & gain & gain err & offset & offset err & $r$ \\
            \hline
            0 & 150.0 & 0.2 & 15997.8 & 2.2 & 0.9995 \\
            1 & 178.02 & 0.03 & 15466.6 & 0.2 & 0.9997 \\
        \end{tabular}
    \end{table*}

    \begin{table*}
        \caption[]{Example of a report on the light output results for the S-mode. The table displays the corresponding channel in column 1, and its light output and error in columns 2 and 3.}
        \label{slo}
    \centering
        \begin{tabular}{|c|c|c|}
            \hline
            channel & light out & light out err \\
            \hline
            0 & 14.72 & 0.04 \\
            1 & 15.29 & 0.01 \\
        \end{tabular}
    \end{table*}

    \begin{table*}
        \caption[]{Example of a report on the spectral resolution for the X-mode. The table displays the corresponding channel in column 1, and the resolution and its error in energy units in the following columns, for each given source.}
        \label{res}
    \centering
        \begin{tabular}{|c|c|c|c|c|c|c|}
            \hline
            source & \multicolumn{2}{c|}{Fe-55 (5.9 keV)} &  \multicolumn{2}{c|}{Cd-109 (22.1 keV)} &  \multicolumn{2}{c|}{Cd109 (24.9 keV)}  \\
            parameter & resolution & resolution err & resolution & resolution err & resolution & resolution err \\
            \hline
            channel			& & & & & & \\
            0 & 0.300 & 0.007 & 0.430 & 0.007 & 0.45 & 0.02 \\
            1 & 0.352 & 0.005 & 0.476 & 0.009 & 0.49 & 0.03 \\
        \end{tabular}
    \end{table*}

\subsubsection{Plots}
\label{subsec322}

The \mescal\ pipeline produces several plots displaying the main performance parameters of the detector. Most of the plots are self-explaining: an uncalibrated, raw spectrum per channel, the calibrated spectrum as seen by each separate channel (for both X and S modes), and gain, offset and light output per channel, including errors and 25-75 percentile levels. However, there are some plot products that are worth explaining in detail.

\begin{figure}
    \centering
    \includegraphics[width=0.5\textwidth]{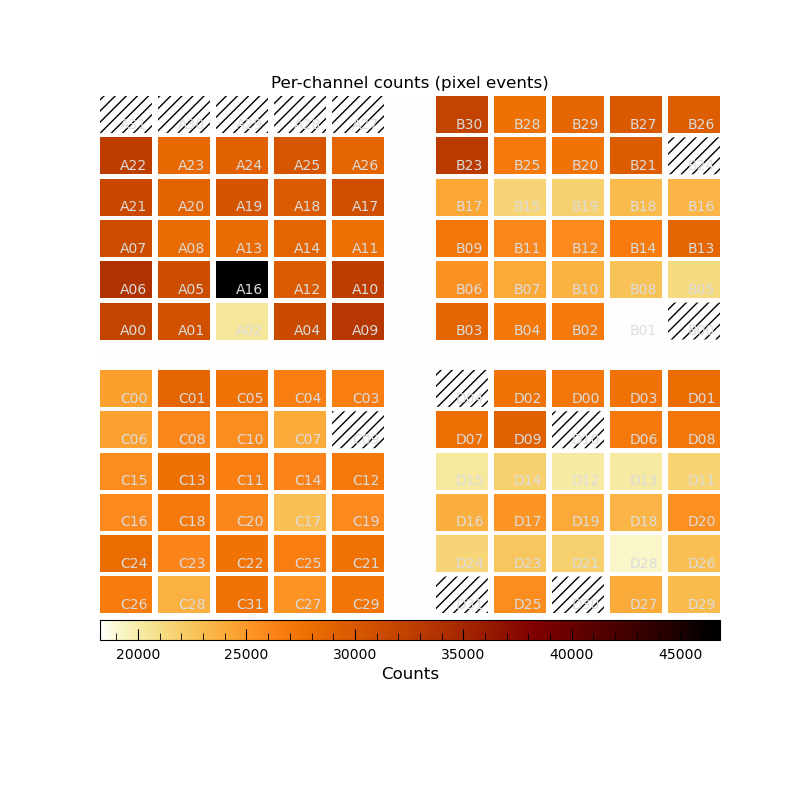}\\
    \caption{The \mescal\ counts map. The grid represents the physical position of each channel in the detector plane, and the color code indicates the total number of detected events per channel. Dashed channels are those momentarily switched off for several reasons (e.g., noise).}
    \label{fig:mapcounts}
\end{figure}

As a first check on whether the data acquisition was performed properly, \mescal\ generates a color-coded counts map (see Figure~\ref{fig:mapcounts}). This map shows each channel as a square in a grid, ordered by their actual position in the detector and tagged with their nominal front-end channel address. The color code shows, by intensity, how many events were detected by each channel during the observation which is being analyzed. Thus, it can easily show if there are any malfunctioning pixels suffering from saturation or retrigger issues, where these pixels are positioned within the detector plane, and to which channels are they coupled to for the S-mode detection (i.e., which is the scintillator crystal that they are reading out). By default, the map does not distinguish between X- and S-mode events. Saturated pixels will be discarded automatically later on by \mescal\, since these pixels do not show any emission lines. Pixels suffering from retrigger are harder to assess, since they may show real emission lines together with artifacts generated by noise. In these cases, \mescal\ discards pixels in which no calibration function can be found.

The diagnostic plots show the uncalibrated, but mode-separated spectra. These plots are generated only for the channels which are switched on during the data acquisition, and only if the peak detection algorithm has found an adequate number of features. Thus, failure to build a diagnostic plot for a certain channel means it is either turned off, saturated, or non-functional. The latter two occurrences are reported in a log file as warnings.

\begin{figure}
    \centering
    \includegraphics[width=0.5\textwidth]{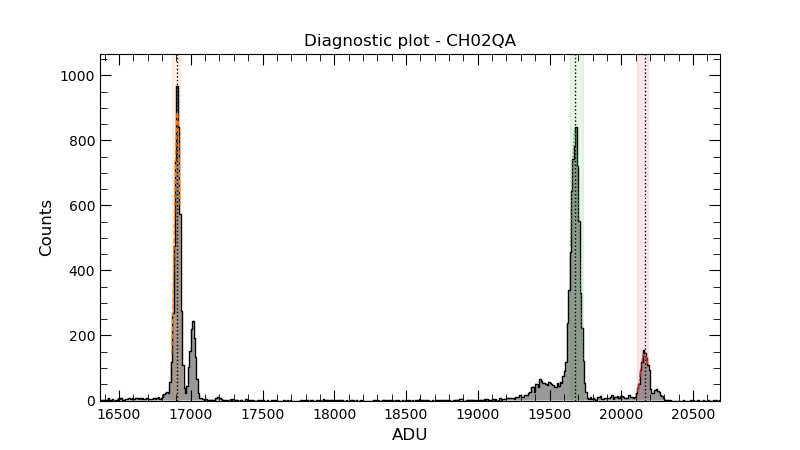}\\
    \caption{The \mescal\ X-mode diagnostic plot for a given channel. The gray histogram is built with the raw, filtered data in instrumental units. The shaded areas indicate the ranges, in instrumental units, in which \mescal\ automatically identifies that a calibration line is present and will be fitted. The dashed color lines represent the centroid line position obtained with such a fit. In this case, the radioactive sources employed are $^{55}$Fe and $^{109}$Cd.}
    \label{diag}
\end{figure}

When a diagnostic plot is built, it shows the ranges where the peak detection algorithm has found emission features to be fitted. Moreover, in the corresponding solid color, it shows the actual Gaussian profile fit over the raw data. Since the peak detection and the fitting are performed by different algorithms, any mismatch between them or with the raw data is well apparent from this plot. Thus, it can provide hints on whether the fitting parameters or the number of emission lines to fit provided by the user were correct or not. This procedure is done both for the X- and S-modes, even if the S-mode usually includes only one emission line as reference. An example of the X-mode diagnostic plot is shown in Figure~\ref{diag}.

\begin{figure}
    \centering
    \includegraphics[width=0.5\textwidth]{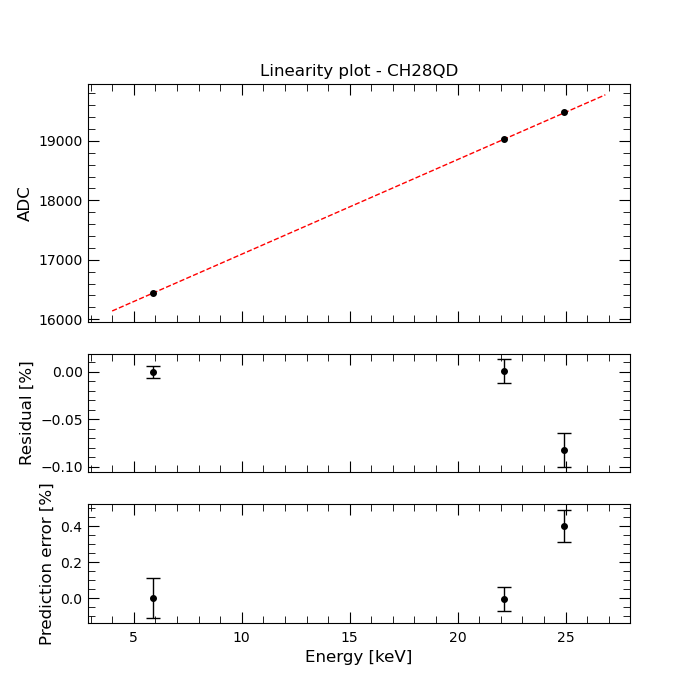}\\
    \caption{The \mescal\ X-mode calibration function linearity plot for a given channel. The upper panel shows the linear least squares best fit to the data, with errors, and the central and lower panel show the residual of the fit and the prediction error, in percentiles, respectively.}
    \label{lin}
\end{figure}

Among the several plots pertaining to each individual channel, \mescal\ produces a \emph{linearity} plot, showing the linear least-squares best fit to the data, the residuals between the fit prediction and the data, and the prediction error, defined as the residual error (see Figure~\ref{lin}). Note that these plots show instrumental units as a function of energy units, since the fit is performed over the inverse of the calibration function (see Section~\ref{subsec31}). The LYRA ASIC is known to be linear to within $\pm$1\% \citep{gandola21}, thus it is expected that the data are fitted with a similar degree of accuracy.

\begin{figure}[ht]
    \centering
    \includegraphics[width=0.5\textwidth]{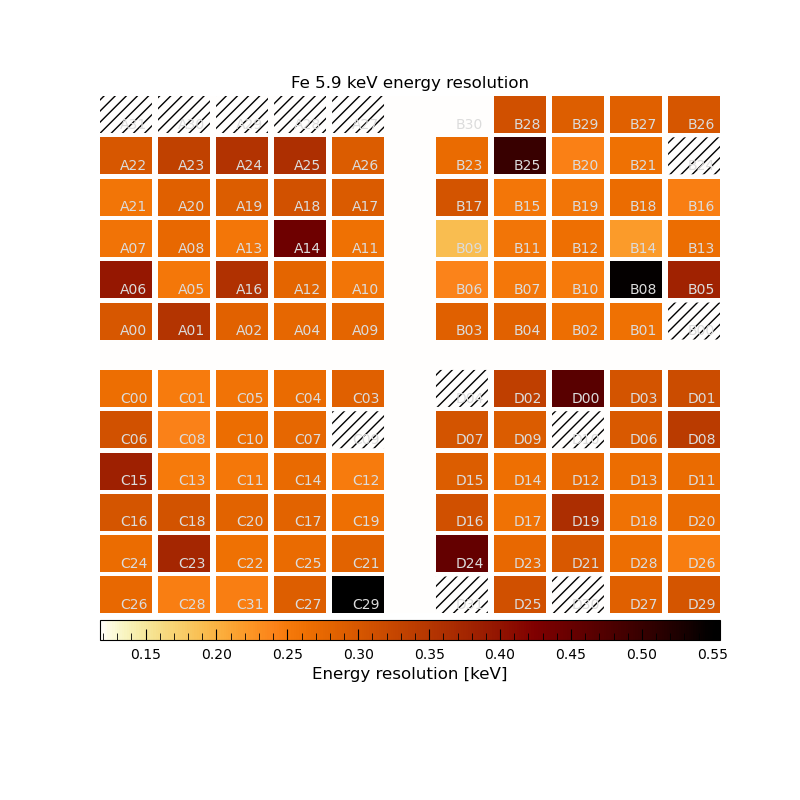}\\
    \caption{The \mescal\ X-mode resolution map for the Fe 5.9 keV emission line. The grid represents the physical position of each channel in the detector plane, and the color code indicates the energy resolution in keV for each pixel.}
    \label{resmap}
\end{figure}

As a by-product of the calibration pipeline, \mescal\ provides a resolution map, taking as reference value the FWHM of the calibration line at the lowest energy. Using the same scheme as the counts map, the resolution map shows each pixel in its position within the detector plane, and their resolution in a color scale. Thus, any evidence of electronic noise issues may be evident in this map (see Figure~\ref{resmap}).

\begin{figure}[ht]
    \centering
    \includegraphics[width=0.5\textwidth]{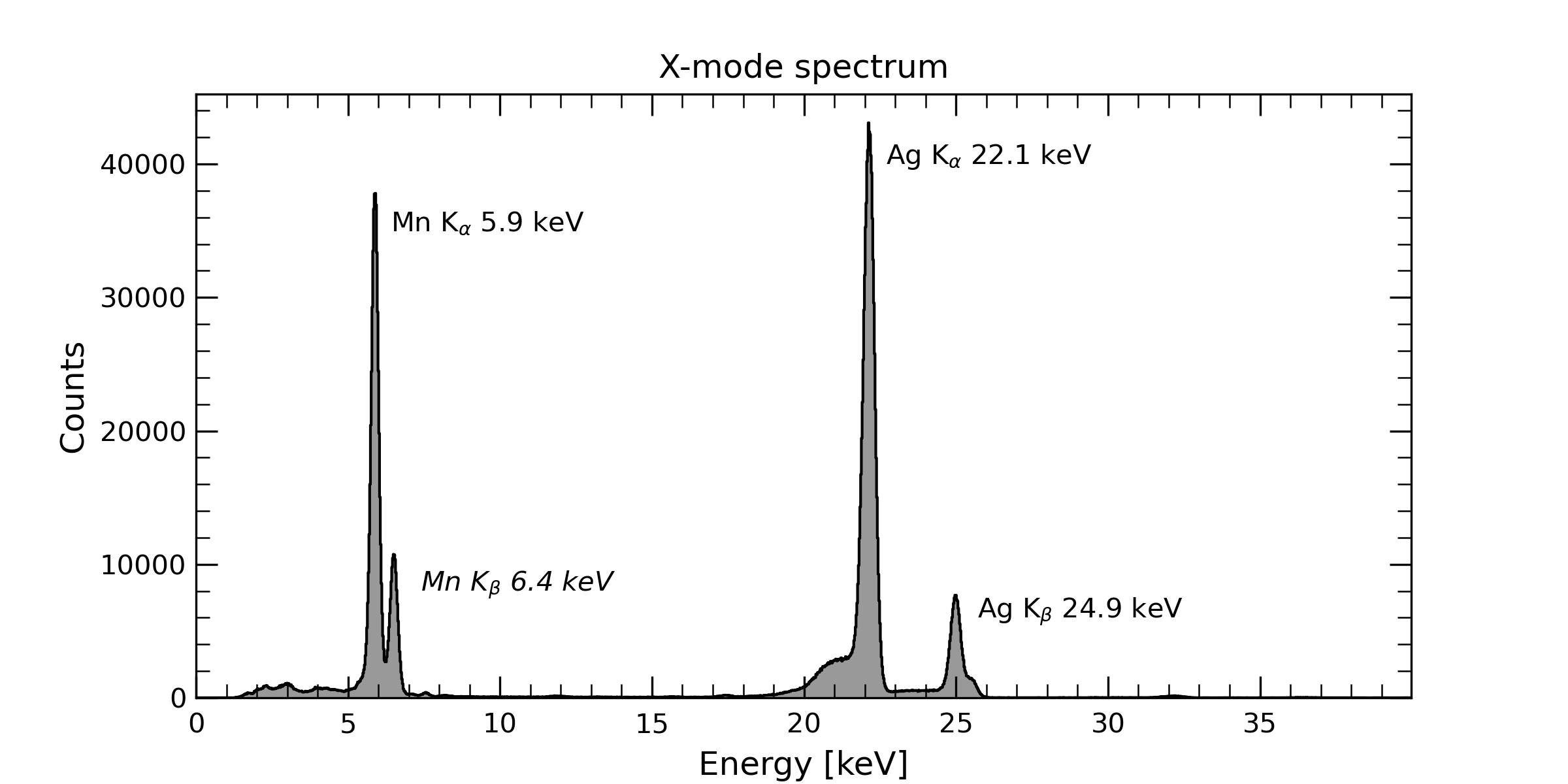}\\
    \caption{The X-mode spectrum of a sample of calibration sources, as seen by the whole HERMES detector. Emission lines in italic were not used for calibration.}
    \label{specx}
\end{figure}

\begin{figure}[ht]
    \centering
    \includegraphics[width=0.5\textwidth]{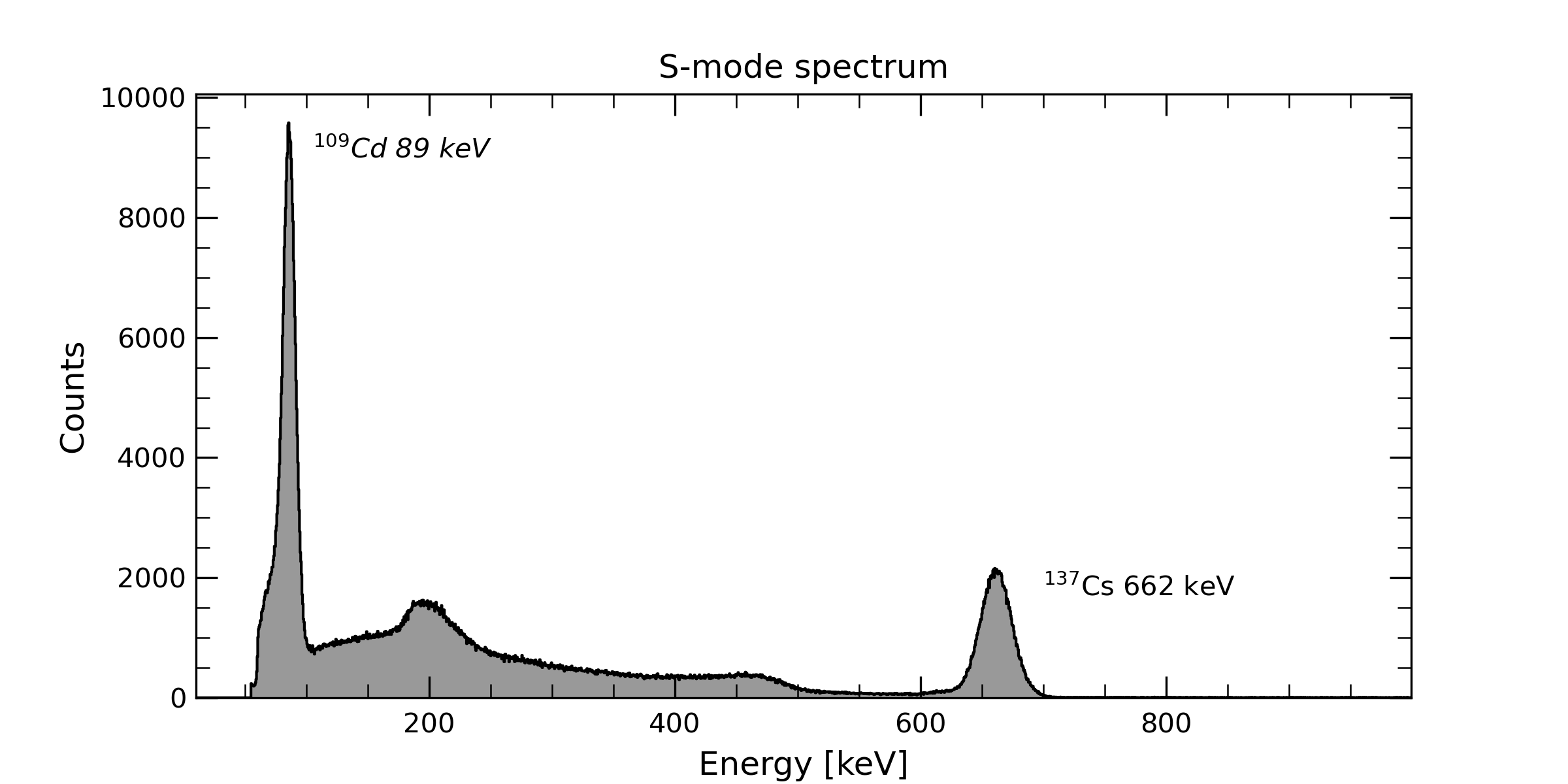}\\
    \caption{The S-mode spectrum of one calibration source, as seen by the whole HERMES detector. Emission lines in italic were not used for calibration. A bump due to Compton scattering and emission lines not used during the calibration procedure is present in the lower energy part of the spectrum.}
    \label{specg}
\end{figure}

Finally, the last output provided by the pipeline are the energy-calibrated X-mode and S-mode spectra. Since each channel works separately, \mescal\ produces one spectrum per mode, per channel. However, since these spectra are all calibrated into energy units, a single spectrum can be obtained by stacking all the calibrated events into one histogram, per detection mode. They represent the overall spectroscopic performance of the whole detector. An example of an X-mode spectrum taken by an HERMES Flight Model can be seen in Figure~\ref{specx}, while an S-mode spectrum can be seen in Figure~\ref{specg}.

\section{Algorithms within \mescal}
\label{sec4}

The \mescal\ pipeline includes several algorithms to perform the energy calibration, data formatting and presentation with the HERMES raw spectra. The majority of this pipeline profits from the \texttt{pandas}, \texttt{numpy} and \texttt{lmfit} Python packages \citep{harris2020array, reback2020pandas, newville2015non}, versions 1.5, 1.23 and 1.2, respectively.
Most algorithms included within it are standard spectroscopic calibration procedures, such as a least-squares linear fit to relate the centroids of the detected calibration lines, in instrumental units, to their energy equivalents, which yields the energy calibration function (see Section~\ref{subsec31}).

In this section, we comment on the algorithms within \mescal\ that automatically perform tasks related to the specific case of the dual-mode (siswich) spectroscopic calibration. We also discuss the implementation of user interaction with these algorithms.

\subsection{X-ray/$\gamma$-ray event discrimination logic}
\label{subsec41}

The event discrimination process aims to classify the events as ‘S’ or ‘X’ (see Sect. \ref{subsec31}). To do this, it has to determine whether a given \emph{multiple} event (i.e., when more than one channel simultaneously trigger) was detected by a pair of channels coupled to the same scintillator crystal, or by independent channels. Thus, the location of the scintillator crystals and their coupling to each pixel must be given as input to correctly identify each event.

Each uncalibrated event detected by the FEE is represented as a row in the raw event list stored in the Level 0/0.5 file. The columns of the event list contain the event arrival time and amplitude, and an address for the channel from which the event was collected. 

Prior to tagging, the event list contains no explicit information on whether it was detected due to the scintillation light induced by the absorption of a $\gamma$-ray photon in the crystal, or due to the direct absorption of an X-ray photon in the silicon bulk of the SDD. After tagging, \mescal\ adds to the event list a new column, containing a string flag which specifies whether the event is an X-mode or an S-mode event.

The logic behind event tagging is straightforward: entries in the event list are aggregated according to their arrival times; then groups of simultaneous events are scanned, looking for events detected on SDD channels coupled to the same scintillator crystal. The correctness of this procedure requires the event rate to be low enough for the probability of observing two simultaneous, low-energy photons on two SDDs coupled to the same scintillator to be negligible. This is indeed the case for the photon count rates up to those expected from a very bright burst \citep{campana20}. Moreover, to completely avoid including photons from intra-scintillator Compton scattering, \mescal\ considers events with multiplicity of either 1 or 2. This means that only single-channel events (multiplicity 1) or double-channel events, with both channels coupled to the same scintillator crystal (multiplicity 2) are used for calibration. Internal Compton scattering events, by definition, should be of multiplicity 4 or higher, since such an event is detected by at least two different scintillators. Thus, by considering only multiplicities of 1 or 2, the contamination present in the spectrum of reference for calibration purposes is reduced.

However, a direct implementation of such a logic in Python can result in a computationally inefficient code. To achieve good performance, \mescal\ leverages on the vectorized operations provided by the \texttt{pandas} data analysis library \citep{pandas}, version 1.5.

First, the event list is stored as a \texttt{pandas} dataframe. Then, the event tag column is added to the frame, with its values set equal to the values of the SDD identifier column. A dictionary is created with an identifier of each coupled pair of channels. The tag values are then mapped into a boolean through this dictionary: if a second event with equal arrival time and channel couple identifier exists, it is set to True, otherwise to False. Finally, True values are converted into an `S' tag, while False are converted into `X' tags.

\subsection{Automatic emission line detection}
\label{subsec42}

As with most spectrometers, the energy calibration of the HERMES detector is performed by acquiring spectra for radioactive sources with well-known emission lines. Then, a calibration function is built by comparing the known energies of the emission lines to their amplitude in instrumental units. This comparison allows the definition of calibration functions for each of the 120 channels in the detector. These functions are defined as the linear least-squares best fit to the data, weigthed by their uncertainties.

The calibration parameters may change substantially with temperature \citep{campana22}. Since the HERMES detectors will operate across a range of temperatures, to ensue proper energy measurements during the scientific operations, it is necessary to calibrate the detectors at different temperature points. Furthermore, during the assembly of a single payload unit, the detector may undergo multiple calibration procedures to ensure the success of each assembly step. All things considered, the characterization of each of the six HERMES detectors requires numerous calibration steps to be performed, due to the considerable number of channels, integration and testing stages, and the need to account for different temperatures. This fact renders the automation of the calibration procedure unavoidable.

The most challenging aspect to automate in the calibration process is the detection of the emission line in instrumental units. The channels spectra may be contaminated by spurious peaks arising from electronic noise, or due to the inability to distinguish scintillator events when the companion sensor is turned off. Even in the absence of spurious peaks, identifying the correct emission line can be challenging due to dispersion in the intrinsic parameters of the spectroscopic chain, unusual threshold values, or artifacts in the emission line profiles.

The algorithm we developed to detect emission lines, rather then relying on templates, leverages the information available to the user on the features of the peaks. This information is encapsulated in a number of \emph{score functions}, which are used to rank all possible combinations of candidate peaks across different metrics. Ultimately, the combination of peaks with the highest ranking across different metrics is selected as the `winning' combination. This technique can be considered an intuitive solution to a multi-objective optimization problem. As such, it can be applied, in principle, to other scenarios in which best candidates must be selected from a pool of elements, yet defining an unique, meaningful loss function is challenging or impossible.\\

To look for these emission lines in a given histogram of X-mode events, in instrumental units, \mescal\ first finds all the local maxima in the histogram within user-defined parameters, such as height, width and prominence (i.e., the distance between peak and baseline), using the \texttt{findpeaks} function of the \texttt{scipy} package, version 1.11. These parameters define the signal-to-noise ratio (S/N) needed to be able to properly detect the emission lines and perform the calibration. They are set to an initial standard value and are then modified iteratively until the specified number of lines is found. We note that \mescal\ has been designed to derive calibration values in a laboratory setup, thus  the implicit assumption is that the S/N of the spectra is always high enough to properly detect emission lines. If this were not the case, then the aforementioned parameters should be changed accordingly to better reflect the S/N of the acquired data.

All of these local maxima are considered emission line candidates, yet they may also originate from artifacts or non-relevant emission features. If the number $K$ of candidates exceeds the number $N$ of calibration lines to be identified, all the $N!/(K!(N-K)!)$ unique combinations of candidate peaks are stored in a list of $K$-tuples. To mitigate any bias stemming from the ordering of the lines, the list of combinations is shuffled.
Otherwise, if the number of candidate peaks is less than the number of searched emission lines, an error is thrown and the calibration process proceeds with the next sensor.
Note that, since the peak finder algorithm sorts every candidate line in instrumental units, then by definition the candidates are sorted in energy units as well. The list of possible sets is shuffled to ensure minimum bias during the selection process.

To determine the correct identification of the emission lines among candidate peaks, \mescal\ assigns a score (the higher, the better) to each combination of candidate peaks, for a number metrics. These metrics include the sum of the absolute errors relative to an a priori guess on gain and offset, linearity, baseline distance, prominence and width.
Once all the scores have been computed, a unique integer between $1$ and $K$ is assigned to each distinct combination of candidate peaks, for each metric. This integer determines the rank of a particular set, according to each of the metrics. For example, the set with the lowest linearity score is associated with a linearity rank of $1$, while the combination with the highest linearity score is assigned a linearity rank of $K$.
Subsequently, the rankings obtained from different metrics are summed, and the set of candidate peaks with the highest overall rank is selected as the best set. To solve ties, if any, the combination with the best ranking for a single metric is chosen. Alternative methods for resolving ties are easy to conceive, such as establishing a hierarchy of scores.
However, opting for ranks has proven to result in more reliable and robust performances. Rankings provides a mechanism to mitigate the impact of possible outliers in scoring and allows for more loosely defined score functions. We estimate the success rate in the automatic identification of emission lines to be greater than 97\%, over a sample of $\sim$10000 calibrated spectra. The few cases in which the identification fails, it is due to instrumental artifacts generated by either the test equipment or by the pixel itself being faulty.

While the algorithm has demonstrated a high success rate in identifying the emission line peaks, there is no inherent guarantee that the combination with the highest rankings will always correspond to the true emission lines from the calibration source. To address potential inaccuracies in line identification, \mescal\ provides users with the option to manually define the emission line boundaries for any channel through a command line interpreter. 

\begin{figure*}
    \centering
    \includegraphics[scale=0.50]{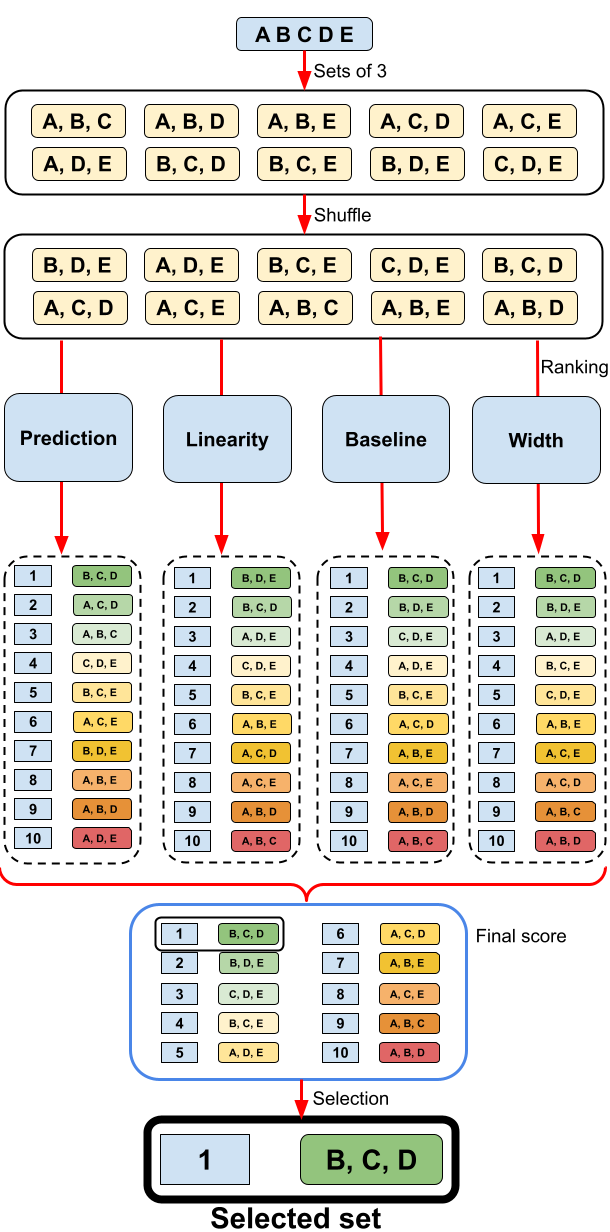}\\
    \caption{Flux diagram explaining the line identification process for a simulated example. If a set of 5 local maxima (labeled A, B, C, D, and E, in increasing order of energy) are found in a given X-mode histogram, and a number of three calibration lines are given as input, then the algorithm first builds all possible combinations. Then, it proceeds to rank every set according to the different criteria, and finally picks the set ranking best across different metrics, which in this case is represented by the set B, C, D.}
    \label{ident}
\end{figure*}

A flux diagram summarizing the line identification algorithm is shown in Figure~\ref{ident}. The scoring metrics we defined are:
\paragraph{A priori error score} For each peak, \mescal\ computes the distance in instrumental units between the peak and an estimated peak position derived from an informed guess of the sensor average gain and offset. The sum of the distances for all peaks in a candidate line set is multiplied by one to yield the a priori error score associated with that combination.\\

\paragraph{Linearity} A preliminary linear fit of the peaks in a combination is performed through a least-squares linear regression (see Sect. \ref{subsec21}). In this run, no initial guess is given. The squared correlation coefficient $r^2$ for each linear function is derived and used as the linearity score for that combination.

\paragraph{Baseline distance} Similarly, the pipeline ranks every set according to the position of the minimum recorded amplitude value. In instrumental units, it consists of a non-zero value which depends of several aspects of the hardware configuration. Different channels do not necessarily share the same instrumental value. However, in energy units and for reasonable values of the threshold, it is expected that the corresponding threshold energy should be close to 2 keV, although the precise value may fluctuate between channels. This assumption arises both from the requirement on the operating band, and from experience: over 12000 calibrated spectra have shown that variations on this value are small. 
Using the preliminary gain and offset parameters obtained when assessing the linearity scores, \mescal\ associates to a given combination of candidate peaks a score equal to the opposite of the distance of the minimum energy value (i.e., the first non-zero bin) to the expected 2 keV.

\paragraph{Peaks prominence} The peak prominence score is defined as the product of the heights of the peaks in a given combination.

\paragraph{Features width} Finally, the last ranking is based on the dispersion of the FWHM of all the lines; since the detected width should be constant for all lines and due only to the overall spectral resolution of the system, (once the Fano noise has been subtracted, \citealt{perotti99}). Artifacts and noise-induced lines, if any, are not bound by this restriction, thus the less dispersion is present in the FWHM of a set of lines, the higher the ``width'' ranking.

\subsection{User interaction}
\label{subsec43}

The \mescal\ pipeline prompts the user through a command line interpreter after the first automatic calibration has been completed. Through this tool, the user is given access to a number of option to display data and diagnostic tools, export optional data products, manually tune calibration of individual channels and re-launch the calibration. The command line interpreter was developed using a customized implementation of Python standard library \texttt{cmd} module and employs elements from the \texttt{beaupy} library. The terminal user interface of \mescal, including the command line interpreter, utilizes the \texttt{rich} Python library, version 12.5, to format console output.

To use this tool, the user is presented with a prompt menu, listing several options, ranging from the output file format (xls, csv or fits), to the immediate display of the different parameter maps (see Sec. \ref{subsec322}). Two of such options, \textsc{setxlim} and \textsc{setslim}, allow the user to display the X-mode and S-mode spectra, respectively; as an interactive window where the user can zoom in or out. The user can, after exploring these spectra, manually insert the X-axis range, in instrumental units, in which \mescal\ should find each specific calibration line. After inserting all required values, \mescal\ can be re-run to perform a new calibration with these ranges as inputs, facilitating the proper line identification. This manual approach, has proven to be useful in the cases in which the automatic line identification fails, due to the presence of statistical artifacts and/or excessive noise.

\section{Summary and conclusions}
\label{conc}

The HERMES calibration pipeline, \mescal, is a software developed specifically for the HERMES raw data calibration and formatting. It is the first pipeline prepared for the calibration of \emph{siswich} detectors. This tool provides a highly automatized solution for the energy calibration of hundreds of spectra, including several instrumental diagnostic tests and statistical checks. 

In particular, the \mescal\ pipeline is capable of generating spectra while tagging events as X-mode or S-mode, following the detection logic of a \emph{siswich} instrument. It can then provide spectroscopic calibration through the automatic identification of calibration lines, the accuracy of which is based on the ranking score of several calibration criteria.

The products of the \mescal\ pipeline are all the necessary calibration parameters, mainly gain, offset and effective light output per channel. These were obtained for different operating temperatures and threshold values, and will be stored in a standardized format as part of the HERMES CALDB.

\mescal\ also includes the possibility of user intervention, if necessary, although it was designed for complete automation of the process. 

We expect to include, once the HERMES models are in orbit, extra functionalities such as flux and time calibration. The \mescal\ structure, supported by python programming, is highly flexible: new algorithms can be included with little to no effort.

\section*{Code availability statement}
\mescal\ is an open-source project released under MIT license. The latest version of the software is  available online at the github repository \url{https://github.com/peppedilillo/mescal}. This paper refers to the version 1.0 of the software, see \cite{giuseppe_dilillo_2023_10065966}.

\section*{Acknowledgements}
This work has been carried out in the framework of the HERMES-TP and HERMES-SP collaborations. We acknowledge support from the European Union Horizon 2020 Research and Innovation Framework Programme under grant agreement HERMES-Scientific Pathfinder n. 821896 and from ASI-INAF Accordo Attuativo HERMES Technologic Pathfinder n. 2018-10-H.1-2020. E. J. M. would like to thank Dr. R. I. P\'aez, for her useful feedback. E.J.M. acknowledges support from the MiniGrant INAF RSN5 research grant ``SISCO'', n. C33C22001040005. 
We wish to thank the ``Summer School for Astrostatistics in Crete'' for providing training on the statistical methods adopted in this work.




\bibliographystyle{elsarticle-harv} 
\bibliography{example}





\end{document}